\documentclass{PoS}

\usepackage{amsmath}

\title{Vector boson (W,Z) Studies with CMS}

\ShortTitle{W, Z boson at CMS}

\author{\speaker{Jiyeon Han}\\
        University of Rochester, NY, USA\\
        E-mail: \email{jyhan@fnal.gov}}

\abstract{We report on recent measurements of W and Z boson production at CMS.
These include measurements of inclusive and differential  W and Z boson production cross sections  at  $\sqrt{s}$ = 8 and 13 TeV,    measurements of the  angular coefficients of Z boson production, forward-backward asymmetry of Z/$\gamma^{*}$ events, and  W charge asymmetry at $\sqrt{s}$ = 8 TeV. These measurements can be used to improve PDFs and refine QCD calculations of weak boson production processes. In addition,  a W-like mass measurement using Z boson events has been  performed to probe the experimental uncertainties  expected for a W mass measurement at $\sqrt{s}$ = 7 TeV.    }

\FullConference{XXIV International Workshop on Deep-Inelastic Scattering and Related Subjects\\
		11-15 April, 2016\\
		DESY Hamburg, Germany}

\begin{document}
\section{Introduction}
   W and Z boson production processes in hadron collider are well understood and have unique signatures and high rates.  
  Inclusive and differential  measurements  of W and Z boson production cross section  can be used to refine calculations based on higher-order  perturbative quantum chromodynamics (QCD) and improve determinations of  parton distribution functions (PDFs) in global PDFs fits.  Here, we present the latest measurements of  vector boson production in pp collisions  with the CMS detector
  at the Large Hadron Collider (LHC)  at $\sqrt{s}$ = 7 and 8 TeV (Run I) and $\sqrt{s}$ = 13 TeV (Run II).
\section{CMS Run II W or Z measurements at $\sqrt{s}$ = 13 TeV}
The inclusive W and Z boson production cross sections are measured with a sample corresponding to $\int{L}$ = 43 $pb^{-1}$ at $\sqrt{s}$ = 13 TeV \cite{inc_wz}.
Both electrons and muons are used to reconstruct W and Z boson with  the transverse energy $\rm{E_{T}}$>25 GeV and $|\eta|$<2.5 for electrons, and  $\rm{p_{T}}$>25 GeV and $|\eta|$<2.4 for muons, respectively.
Identification and selection criteria for electron and muon candidates are applied and energy isolation is required.
To extract the W boson signal, the missing transverse energy distribution is fit to a sum of three components: the W boson signal template, the QCD background template, and other backgrounds template  (e.g. $W\to\tau\nu$, Drell--Yan, diboson, and top-pair production).  The background contamination in the Z boson sample is relatively small compared to the W boson case.
 All  background processes for Z boson candidates are estimated
 with MC simulations in the mass range of 60<M(Z)<120 GeV.
The overall systematic uncertainties in the cross sections are 2.5\% for the electron and 2.0\% for the muon channels, respectively. 
In addition, there is an overall luminosity normalization uncertainty of 4.8\%.  The uncertainty in the theoretical prediction is 2\%.
The results for the electron and muon channels are combined  and 
compared with the theory prediction.
The measured inclusive cross section is 19950$\pm$70(stat.)$\pm$360(syst.)$\pm$960(lumi.) pb for W$\to \ell\nu$  and
 1910$\pm$10(stat.)$\pm$40(syst.)$\pm$90(lumi.) pb for Z$\to \ell^{+}\ell^{-}$, respectively. These  are in a good agreement with FEWZ predictions in  next-to-next-to-leading order (NNLO QCD with NNPDF3.0 PDFs) which are
  $\sigma(\rm{W} \to \ell\nu)$=19700$\pm$520 pb and  $\sigma(\rm{Z} \to \ell^{+}\ell^{-})$=1870$\pm$50 pb, respectively.
In addition, the inclusive cross sections for W$^{+}$ and W$^{-}$ and the  ratio of W to Z boson production cross
 sections \cite{inc_wz} have been measured.

\begin{figure}
 \begin{center}
\includegraphics[width=.32\textwidth]{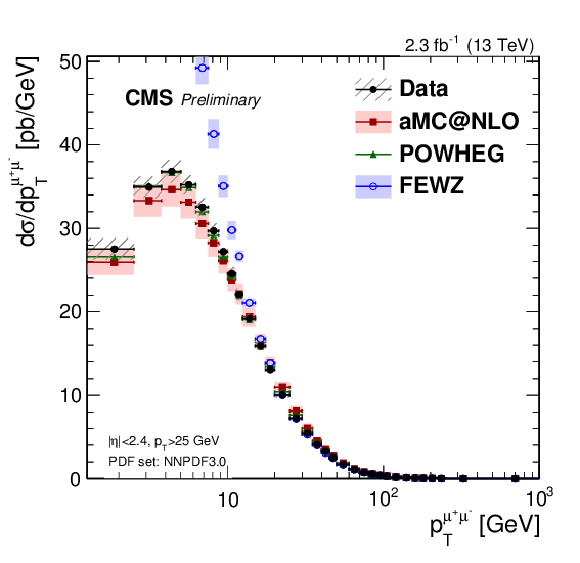}
\includegraphics[width=.32\textwidth]{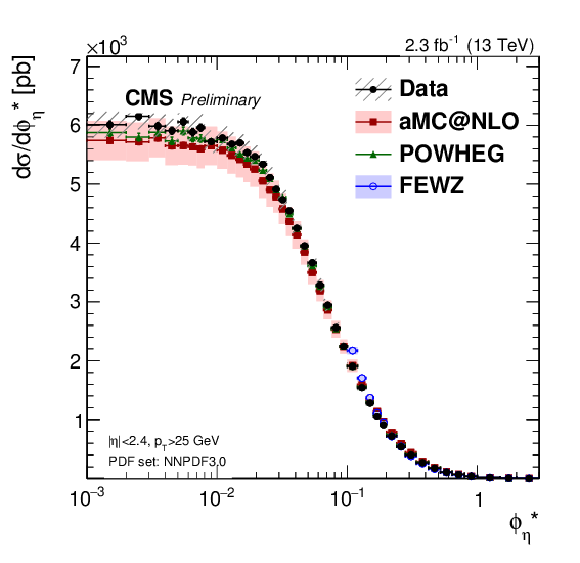}
\includegraphics[width=.32\textwidth]{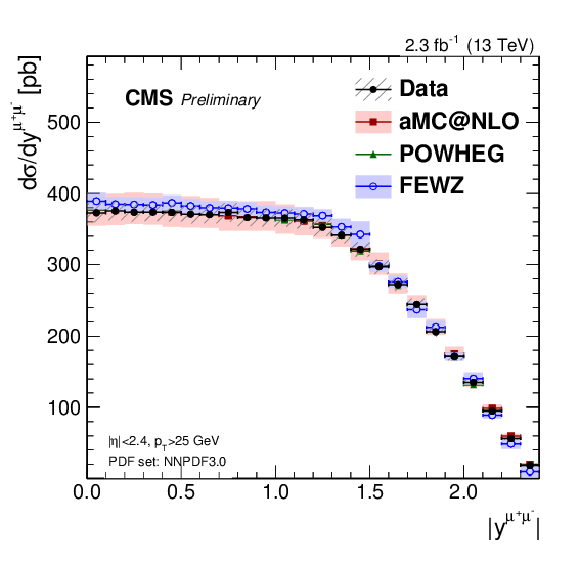}
 \caption{Differential cross sections for Z boson production  as a function of $\rm{P_{T}}$, $\phi_{\eta}^{*}$, and $y$. The measured cross sections are compared to MC@NLO, POWHEG(NLO), and FEWZ(NNLO) theory predictions.} 
  \label{fig3}
 \end{center}
 \end{figure}

The differential cross sections for Z boson production are measured as a function of Z boson transverse momentum ($\rm{P_{T}}$), rapidity ($y$),   and $\phi_{\eta}^{*}(=\tan(\frac{\pi - \Delta \phi}{2}) \cdot \sin(\theta_{\eta}^{*})$) using the full sample at $\sqrt{s}$ = 13 TeV collected in 2015 ($\int{L}$ = 2.3 $fb^{-1}$).
 The differential production cross section  as a function of $\rm{P_{T}}$ tests gluon resummation and parton shower models at low $\rm{P_{T}}$, and perturbative QCD calculations at high $\rm{P_{T}}$ (where the $qg$ scattering process is dominant). 
 The distribution in  $\phi_{\eta}^{*}$ also  probes the Z boson $\rm{P_{T}}$ spectrum.  However, since $\phi_{\eta}^{*}$  only depends on the direction of the leptons, this distribution has a   smaller experimental error. 
  The differential cross section as a function of  $y$ can be used to constrain PDFs in global PDF fits. 
  Only muons are used for the differential cross section measurements.  Here, Z boson candidates are selected with  $\rm{p_{T}}(\mu)$>25 GeV,  $|\eta|$<2.4, and an dimuon invariance mass 60<M($\mu\mu$)<120 GeV, corresponding to $\sim$1.3 million Z events.
 Corrections are applied for the  muon $\rm{p_{T}}$ scale  (in both data and MC), and efficiency scale factors of data to MC are applied to the MC.    The backgrounds are determined with MC simulations.
 The differential cross section measurements are unfolded within the fiducial volume and the measurements are compared to
 the following theory  predictions:  MC@NLO, POWHEG(NLO), and FEWZ(NNLO) with NNPDF3.0 PDFs.
The  differential cross section measurements as a function of  $\rm{P_{T}}$, $\phi_{\eta}^{*}$, and $y$  are compared with theory predictions in   Figure \ref{fig3}.
  The FEWZ prediction deviates from the data at low $\rm{P_{T}}$. This is expected due to the absence of resummation in FEWZ.
  Otherwise, the measurements are a good agreement with theory predictions within errors.  
 Additional  details of these measurements are given in ref. \cite{dif_z}.
  
\section{W and Z CMS measurements at $\sqrt{s}$ = 8 TeV (Run I)}
The double differential cross section as a function of  Z boson $\rm{P_{T}}$ and $y$ is measured using a sample
corresponding to   $\int{L}$ = 19.7 $fb^{-1}$ at $\sqrt{s}$ = 8 TeV \cite{double_pty}.
 In addition to  muon identification and isolation requirements, the leading (sub-leading) muon is required to have $\rm{p_{T}}(\mu)$>25 (10) GeV and $|\eta|$<2.1 (2.4).
A tight mass range, 81<M($\mu\mu$)<101 GeV, is used to select the Z boson candidates. 
The double differential cross section is unfolded to pre-final-state radiation (FSR) within the muon kinematics. 
Both absolute and normalized differential cross sections ($\frac{d^{2}\sigma}{d\rm{P_{T}}dy}$ and $\frac{1}{\sigma_{inc}}\cdot\frac{d^{2}\sigma}{d\rm{P_{T}}dy}$) are measured \cite{double_pty} and compared to the theory prediction of  FEWZ(NNLO) with NNPDF2.3 PDFs and radiative corrections.  The measured cross sections are in  agreement with theory within errors.

 
 The general structure of the lepton angular distribution in the boson rest frame is given by
\begin{align*}
\frac {d^2 \sigma } {d\cos\theta^{*}d\phi^{*}}
\propto \Bigl[(1+\cos^2\theta^{*}) +A_0 \frac{1}{2}(1-3\cos^2\theta^{*}) + A_1\sin(2\theta^{*})\cos\phi^{*} + A_2\frac{1}{2}\sin^2\theta^{*}\cos(2\phi^{*})    \\
 +A_3\sin\theta^{*}\cos\phi^{*} + A_4\cos\theta^{*} + A_5 \sin^2 \theta^* \sin(2\phi^*)+A_6\sin(2\theta^*)\sin{\phi^*} + A_7\sin{\theta^*}\sin{\phi^*}\Bigr].
\end{align*}
Here, $\theta^*$ and $\phi^*$ are the polar and azimuthal angles of the negatively charged lepton in the rest frame of the lepton pair.
 The angular coefficients ($A_{0}$ to $A_{4}$) of Z boson are measured using the same sample and selection criteria as used
for the double differential cross section measurement as a function of  Z boson $\rm{P_{T}}$ and $y$ at $\sqrt{s}$ = 8 TeV \cite{angular}.
 The angular coefficients, $A_{0,1,2}$, are related to the polarization of the  Z boson and $A_{3,4}$ originate from $\gamma^{*}$/Z interference. Here, $A_{0,2}$ determine the fraction of $q\bar{q}$ and $qg$ processes in pp collisions as a function of
  Z boson $\rm{P_{T}}$. The $A_{4}$ parameter is directly  related to the forward-backward asymmetry of Z boson production ($A_{FB}$)  and  is sensitive to the electroweak mixing angle, $\sin^{2}\theta_{\rm{W}}$.
 The angular coefficients are measured using the template fitting method in Collins-Soper frame \cite{collins-soper} as a function of Z boson $\rm{P_{T}}$ and $|y|$.
 The measured angular coefficients are compared with MadGraph, POWHEG(NLO), and FEWZ(NNLO). There is good agreement with the MadGraph and FEWZ(NNLO) predictions.
 Since the MadGraph calculation doesn't include EW radiative correction, the MadGraph prediction has a  higher $A_{4}$ than  the other theory predictions.
Details of  the measurement of the  angular coefficients as a function of Z boson $\rm{P_{T}}$ for two rapidity bins, $|y|$<1.0 and 1.0<$|y|$<2.1, are given in ref.  \cite{angular}

The forward-backward asymmetry of Z/$\gamma^{*}$ events ($A_{FB}$) is measured using a sample corresponding to $\int{L}$ = 19.7 $fb^{-1}$ at $\sqrt{s}$ = 8 TeV \cite{afb}.
The $A_{FB}$ parameter depends on dilepton mass, quark flavor, and the electroweak mixing angle $\theta_{\rm{W}}$.
 $A_{FB}$ is extracted from the  difference of the number of forward ($\cos\theta^{*}>0$) and backward ($\cos\theta^{*}<0$) events  divided by the sum of events in bins of  invariant mass (M) and $|y|$.
Muons and electrons in the central and endcap regions  are used to select the dilepton events with lepton $\rm{p_{T}}$>20 GeV and $|\eta|$<2.4.
The measurement of $A_{FB}$ is extended up to $|y|$=5 by
including electrons in the forward calorimeter ($\rm{p_{T}}$>20 GeV, 3<$|\eta|$<5) in conjunction with  a central electron ($\rm{p_{T}}$>30 GeV, $|\eta|$<2.4).

The measured $A_{FB}$ is unfolded for resolution, acceptance, efficiencies, and the effects of  FSR. 
The unfolded $A_{FB}$ is combined for muons and electrons up to $|y|$=2.4. 
The POWHEG prediction with $\sin^{2}\theta_{\rm{W}}$ = 0.2312 is in good agreement with the data.
 Measurements of $A_{FB}$ as a function of invariant mass bins in  $|y|$   are compared with  POWHEG predictions
 in ref. \cite{afb}.

\begin{figure}
 \begin{center}
\includegraphics[width=.32\textwidth]{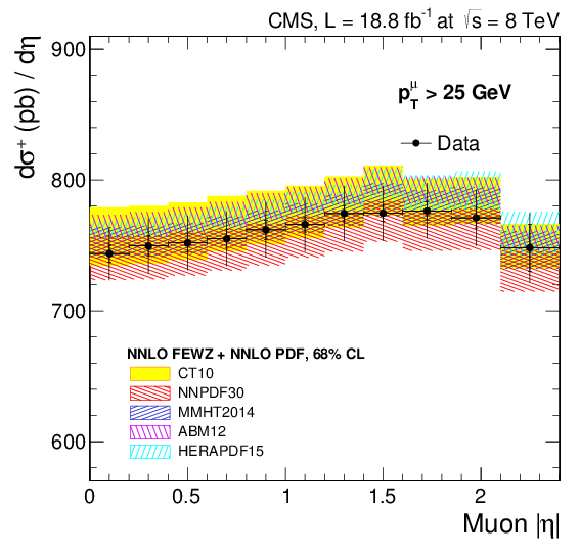}
\includegraphics[width=.32\textwidth]{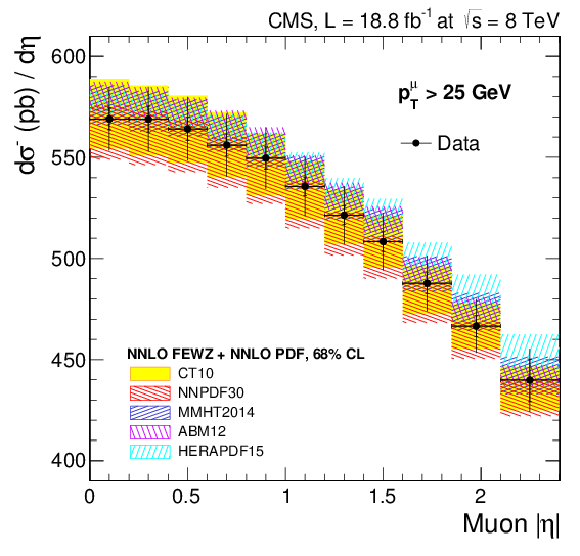}
\includegraphics[width=.32\textwidth]{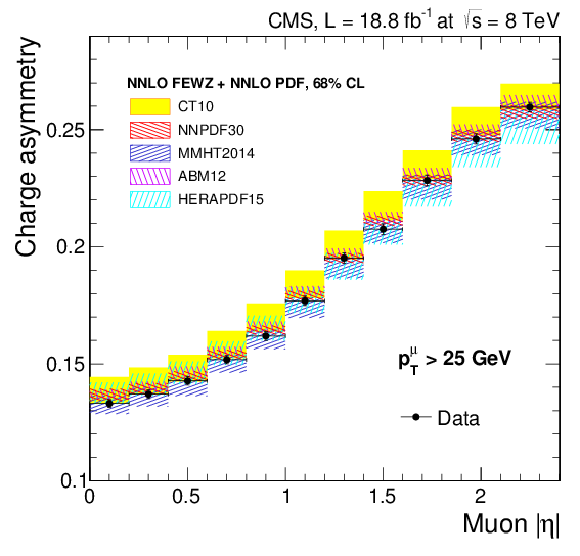}
 \caption{Differential cross sections for W$^{+}$(left), W$^{+}$(middle), and the W charge asymmetry(right) as a function of muon $\eta$.  The measurements are compared to FEWZ predictions with various PDFs sets, (CT10, NNPDF3.0, MMHT2014, HERAPDF1.5 and ABM12).} 
  \label{fig7}
 \end{center}
 \end{figure}

 
 In pp collisions,  the production of W$^{+}$ boson is larger than W$^{-}$ bosons because the  proton is composed primarily of  
  $uud$ quarks. In addition, the W bosons are polarized because of  parity violation, which also  results in an asymmetry in the lepton decay kinematics.
 The charge asymmetry of the W boson decay lepton as a function of $A(\eta)$ is defined as the difference of the cross section 
for W$^{+}$ and W$^{-}$  divided by the sum as a function of  lepton $\eta$.
 The asymmetry  $A(\eta)$ can be used to constrain PDFs,  and in particular the ratio of $u$ and $d$ quark distributions.
 In addition, the differential cross sections of W$^{\pm}$ production as a function of muon $\eta$ ($d\sigma^{\pm}/d\eta$) are measured  
for events with  $\rm{p_{T}}(\mu)$>25 GeV and $|\eta|$<2.4 
using a sample corresponding to  $\int{L}$ = 18.8 $fb^{-1}$ at $\sqrt{s}$ = 8TeV .
 The differential cross sections for W$^{\pm}$ production are corrected for the efficiency and effects of  FSR.
 The measurements of the differential cross sections and  asymmetries are compared to FEWZ predictions with various PDFs models.
  The agreements between data and predictions are within the uncertainties of the PDFs. A QCD analysis is performed at NNLO to test the impact of $A(\eta)$ measurement.
   The $A(\eta)$ measurement significantly improves the determination of the  valence quark PDFs.
  Figure \ref{fig7} shows the measurements of the differential cross sections for W$^{\pm}$ production and  asymmetry as compared to  theory predictions.
Additional details of this measurements  are given in ref.  \cite{wasym}.
 
 \section{W-like measurement of the Z boson mass at $\sqrt{s}$ = 7 TeV}
\begin{figure}
 \begin{center}
\includegraphics[width=.5\textwidth]{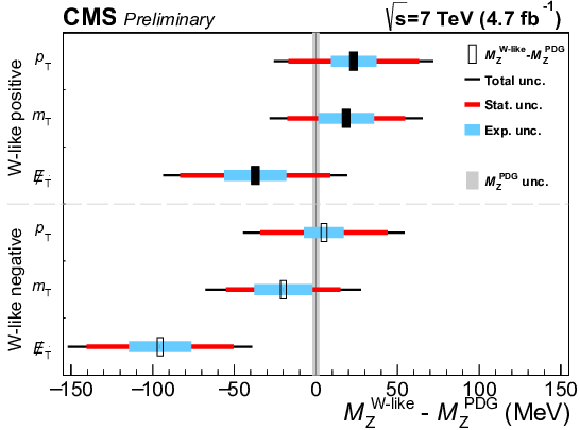}
 \caption{The difference between the fitted W-like mass and $\rm{M_{Z}^{PDG}}$ using each of three observables
with  corresponding statistical and systematic uncertainties. } 
  \label{fig8}
 \end{center}
 \end{figure}

 The W mass ($\rm{M_{W}}$) provides a critical test on the consistency of Standard Model (SM) predictions (using 
 the measured masses of the  top quark ($\rm{M_{t}}$) and  Higgs($\rm{M_{H}}$)).
 The world average of $\rm{M_{W}}$ direct measurements is 80.385$\pm$0.015 GeV.  A global electroweak fit predicts $\rm{M_{W}}$=80.358$\pm$0.008 GeV.
 The world average of $\rm{M_{t}}$ measurements is 174.34$\pm$0.76 GeV and the most precise direct $\rm{M_{t}}$ measurement has an error of 0.66 GeV.  A direct measurement of  $\rm{M_{W}}$  should be at a level of  precision of 6 MeV or better to 
 test the consistency of the SM (given the presently available accuracy of $\rm{M_{t}}$ and $\rm{M_{H}}$).
 A direct W mass measurement is on-going at CMS using Run I data at $\sqrt{s}$ = 7 and 8 TeV.   To test the technique, 
  CMS has measured the Z boson mass by removing one of two muons to form W-like  Z boson candidates.
 This measurement provides a proof of principle and a quantitative validation of  the technique to be used in the measurement of the W mass. 
 The measurement is done with a sample of  0.2 million  Z$\to \mu\mu$ events
corresponding to $\int{L}$ = 4.7 $fb^{-1}$ at $\sqrt{s}$ = 7 TeV. 
The measurement utilizes improvements in calibration of  muon $\rm{p_{T}}$ (<15 MeV) and recoil calibration (<14 MeV).
 Three observables, muon $\rm{p_{T}}$, the transverse mass ($\rm{m_{T}}$), and the missing transverse energy
 are used to extract the W-like mass measurement. 
 The W-like mass measurements from muon $\rm{p_{T}}$,  $\rm{m_{T}}$, and the missing transverse energy fits are compared with  the PDG Z boson mass ($\rm{M_{Z}^{PDG}}$) in Figure \ref{fig8}. 
 The most precise W-like mass measurement is from $\rm{m_{T}}$ fit and it differs from the PDG value by 18 $\sim$ 20 MeV, which is within the uncertainties of the W-like measurement (stat. $\oplus$ syst. = 48 MeV level). Additional  details of the measurement  are given  in ref. \cite{w-like}.
  
\section{Summary}
 We report on recent CMS measurements of W and Z boson production  at $\sqrt{s}$ = 7 and 8 TeV (Run I),  
and  preliminary results for the inclusive and differential cross sections of Z boson production at $\sqrt{s}$ = 13 TeV  (Run II).
 We also report on various  other published CMS measurements with W and Z bosons at $\sqrt{s}$ = 8 TeV.
 These detailed studies of W and Z boson production processes probe various aspects of the SM. In addition, 
 the measurements can be used to provide additional constraints on PDFs and  test QCD  theory predictions.
 A W-like mass  measurement with  Z boson events at $\sqrt{s}$ = 7 TeV provides a cross-check on the technique
 for the W mass measurement  which is on-going.


\end{document}